\documentclass[aip,apl,preprint,longbibliography]{revtex4-1}

\usepackage{graphicx}
\usepackage{amsmath}
\usepackage{amssymb}
\usepackage{natbib}
\usepackage{bm}
\usepackage{color}
\usepackage{siunitx}

\begin{document}
\title{Effects of Marangoni and drag forces on the transition from vapor-rich to air-rich bubbles}

\author{Kyoko Namura}
\email{namura.kyoko.2s@kyoto-u.ac.jp}
\affiliation{Department of Micro Engineering,
Kyoto University,
Kyoto Daigaku-Katsura, Nishikyo-ku, Kyoto 615-8540, Japan}
%

 \author{Takuya Iwasaki}
 \author{Kaoru Nakajima}
 \author{Motofumi Suzuki}

\affiliation{Department of Micro Engineering,
Kyoto University,
Kyoto Daigaku-Katsura, Nishikyo-ku, Kyoto 615-8540, Japan}

\keywords{photothermal, microbubble, Marangoni effect, drag force}

\begin{abstract}
In this study, we investigated the formation of air-rich microbubbles through local photothermal heating of non-degassed water.
When non-degassed water is locally heated, vapor-rich bubbles are initially formed. These bubbles have a maximum radius of approximately 9 \si{\um} and stabilize while oscillating and exhaling air-rich bubbles. However, when the exhaled bubbles fuse and grow, they revert to vapor-rich bubbles on the heat source.
The vapor-rich bubbles are then exposed to a large amount of air, causing them to transition to air-rich bubbles.
In this paper, the motion of the exhaled air-rich bubble is explained by the drag force owing to the flow and the Marangoni force owing to the temperature gradient acting on the bubble.
Because the drag force is proportional to the bubble radius, and the Marangoni force is proportional to the square of the bubble radius, the larger the bubble, the stronger the effect of the Marangoni force.
Thus, as the bubbles grow, they are drawn toward the heat source against the flow created by the vapor-rich bubbles. These results are useful for a better understanding of bubble growth and determining the conditions for the stable formation of vapor-rich bubbles in non-degassed water.
\end{abstract}

\noindent
\textcolor{red}{{\small This article has been submitted to Journal of Applied Physics.
https://pubs.aip.org/aip/jap}}

\maketitle

\section{INTRODUCTION}
Microbubbles have attracted considerable interest for decades as tools for manipulating small volumes of fluids and particles.
The large volume changes of bubbles associated with phase changes and the high compressibility of the gas phase can strongly drive the surrounding fluid and has been used in fluid pumping\cite{Geng2001,Tho2007,Dijkink2008,Patel2014} and mixing,\cite{Liu2002,Hellman2007} printing,\cite{Serra2019} photoacoustic imaging,\cite{Kim2009,Ju2013}, etc.
In addition, the surface tension gradient of a bubble due to a temperature or concentration gradient, known as the Marangoni or thermocapillary force, has been used in the manipulation of bubbles\cite{Ortega2018,Sarabia2021} and particles,\cite{Berry2000,Namura2015,Lee2020,Dara2023} fluid pumping and mixing,\cite{Taylor2004,Jones2020} particle printing,\cite{LinLinhan2016,Fujii2017,Moon2020} manipulation of biological specimens,\cite{Nishimura2014,Tokonami2020} etc.
This wide range of applications has accelerated research on the behavior of bubbles.\cite{Hou2015,Jollans2019,Zeng2021}
Although the applications of fluid control using bubbles vary, particularly in terms of continuous fluid pumping, vapor-rich bubbles generated on a local heat source are expected to be beneficial.\cite{Christopher2005,Li2017,Namura2017}
For example, when degassed water is heated using a heat source with a diameter of approximately 7 \si{\um}, vapor-rich bubbles with diameters of 10--20 \si{\um} are generated at the heat source.\cite{Namura2017,Namura2019,Namura2022}
These bubbles oscillate on the order of sub-MHz, but the time-averaged size of the bubbles remains almost unchanged as long as the heating continues.\cite{Namura2020} The flow created by this bubble exceeds 1 mm/s at a distance of 300 \si{\um} from the bubble and is expected to be on the order of 1 m/s in the vicinity of the bubble. Such flows are expected to be useful for the agitation and pumping of fluids.

A similar phenomenon is known to occur when non-degassed water is heated above a certain heating density threshold.\cite{Li2017,Nguyen2019_2,Nguyen2019}
Here, vapor-rich bubbles 20--40 \si{\um} in diameter are generated above the local heating point and oscillate in the sub-MHz range.
The bubbles continuously absorb dissolved air in the water and exhale it as air-rich bubbles each time they oscillate.\cite{Li2017}
The exhaled air-rich bubbles are transported away from the heat source by the flow created by the vapor-rich bubbles on the heat source.
Consequently, oscillating vapor-rich bubbles, which create a strong flow, remain on the heat source without transitioning to air-rich bubbles.
However, when the heating density of the heater falls below the aforementioned threshold, the bubbles cannot expel the air they have absorbed and eventually stabilize as air-rich bubbles.
Reports show that air-rich bubbles cannot generate convection as strongly as vapor-rich bubbles do, which is expected to be avoided in terms of continuous fluid pumping.\cite{Namura2017}
The process of their formation must be fully understood to avoid the formation of air-rich bubbles.
The formation process and behavior of air-rich bubbles on localized heaters have been extensively investigated by Wang et al.,\cite{WangY2017,WangY2018}  Li et al.,\cite{LiXiaolai2019} and Zaytsev et al.\cite{Zaytsev2020}
They reported that vapor-rich bubbles form first, followed by a gradual transition to air-rich bubbles.
First, giant vapor bubbles are formed, followed by relatively small oscillating vapor-rich bubbles that are retained.
These bubbles gradually absorb air on a timescale of tens of \si{\us} and eventually stop oscillating.
To date, this air uptake has been discussed in terms of the air uptake associated with water evaporation and the diffusion of air molecules through the bubble interface.
These arguments are highly informative and agree well with experimental results.
Recently, we found that in addition to the above-mentioned diffusion- and evaporation-based transitions from water vapor-rich to air-rich bubbles, transitions also occur when water vapor-rich bubbles reabsorb exhaled air-rich bubbles.
Interestingly, the exhaled air-rich bubbles approach the water vapor-rich bubbles against the flow created by the water vapor-rich bubbles under local heating.
Interesting behaviors of air-rich bubbles near the local heating point have been reported in recent years.\cite{Ortega2018,Sarabia2020,Sarabia2021,Zeng2021}
In particular, Sarabia-Alonso et al. successfully manipulated air-rich bubbles near intermittent local heating points using pulsed laser irradiation and discussed the mechanism in detail.\cite{Sarabia2020}
However, vapor-rich bubbles produced by continuous heating are known to generate strong convection currents, and the mechanism by which air-rich bubbles approach vapor-rich bubbles against their flow is not fully understood.

In this paper, we discuss the phenomenon by which vapor-rich bubbles reabsorb exhaled air-rich bubbles.
A continuous wave (CW) laser was focused on a FeSi$_2$ thin film to locally heated non-degassed water via photothermal conversion.
We then observed the motion of the air-rich bubbles exhaled from vapor-rich bubbles during local heating.
By measuring the distance from the heat source to the air-rich bubbles, bubble size, and bubble velocity, we showed that the behavior of the bubbles can be understood using the balance between the drag and Marangoni forces acting on the bubbles.

\section{EXPERIMENTS}
In this study, amorphous FeSi$_2$ thin films were used for photothermal conversion. The details of the thin-film preparation procedure can be found in our previous paper.\cite{Namura2020} A brief description is given here.
An FeSi$_2$ thin film was prepared on a glass substrate using radio frequency magnetron sputtering. An FeSi$_2$ target (purity: 99.99\%) was used as the sputtering source. The deposition chamber was evacuated to a vacuum greater than 1 $\times$ 10$^{-4}$ Pa. Subsequently, Ar was introduced as the sputtering gas at a rate of 9 sccm. The sputtering was performed at room temperature. During sputtering, the total gas pressure was maintained at 0.8 Pa and the substrate was rotated rapidly. Subsequently, an FeSi$_2$ layer of 50 nm was deposited on the glass substrate.

The prepared FeSi$_2$ thin-film chip (8 mm $\times$ 25 mm) was fixed inside a glass cell (10 mm $\times$ 10 mm $\times$ 58 mm, F15-G-10, GL Science) using a pair of magnets. The glass cell was filled with non-degassed ultrapure water (18.2 M\si{\ohm} cm, Millipore-Direct Q UV3, Merck). The prepared fluid cells were then placed in an optical setup equipped with a CW laser source for photothermal heating and an optical microscope for bubble observation, as described in our previous paper.\cite{Namura2020} To generate microbubbles in the non-degassed water, we focused the CW laser with a wavelength of 830 nm (FPL830S, Thorlabs) onto the FeSi$_2$ thin film from the rear side of the glass substrate. The laser spot radius ($1/e^2$) on the film was 3.4 \si{\um}, and the laser power at the sample surface was 115 mW. The bubbles generated on the laser spot were observed from a direction almost parallel to the substrate surface using an objective lens (20 $\times$, NA = 0.40). The observation plane of the microscope was perpendicular to the direction of gravity. Bubble images were recorded using a high-speed camera (FASTCAM Mini AX100, Photron) under green light illumination. A short-pass filter was placed in front of the camera to eliminate the 830 nm laser source. The camera frame rate was set to 30,000 fps.

\section{RESULTS AND DISCUSSION}
\begin{figure}[tbp]
\centerline{\includegraphics[bb = 0 0 896.95 276, width=13cm]{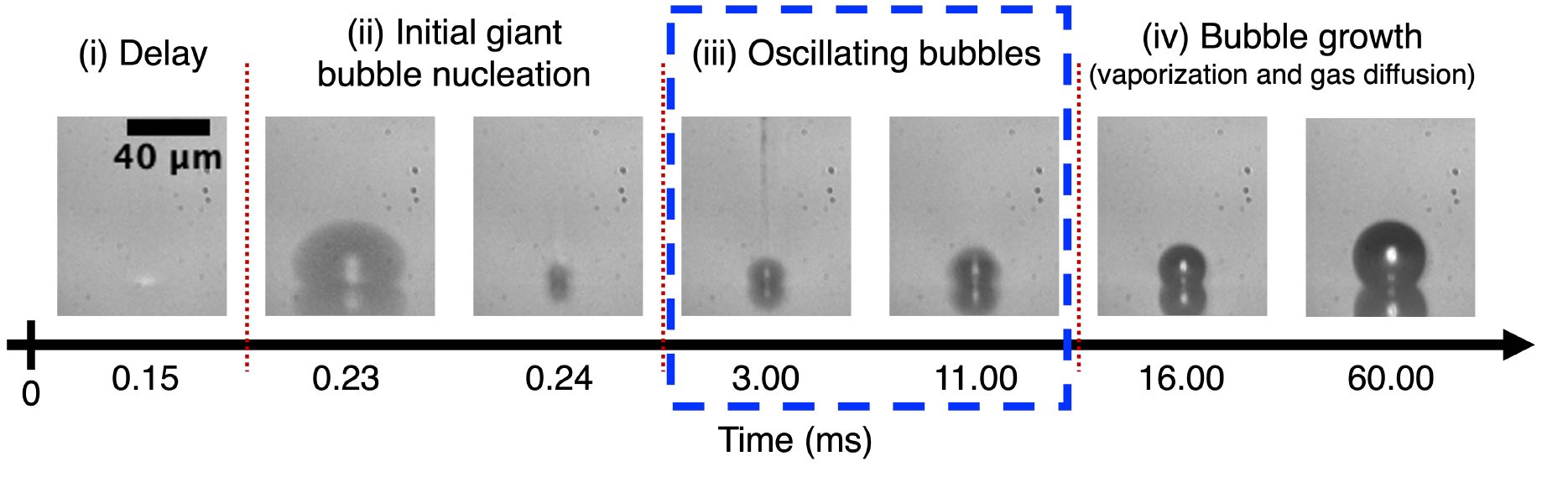}}
\caption{Typical time-series images of bubble formation and growth in non-degassed water. The early stage of bubble formation observed in this study was consistent with that reported by Wang et al.\cite{WangY2018}}
 \label{fig1}
\end{figure}

Figure \ref{fig1} shows typical time-series images of bubble formation and growth in non-degassed water under photothermal localized heating.
The horizontal axis represents the time after the start of laser irradiation.
Phase (i) represents the waiting time after the start of laser irradiation until the fluid reaches the temperature at which bubbles are generated. Subsequently, relatively large and explosive vapor bubbles are generated in phase (ii). 
This bubble disappeared in a short period (less than 10 \si{\us}), leaving behind an oscillating bubble approximately 20 \si{\um} in diameter (phase (iii)). 
These bubbles grow through the evaporation of water and uptake of dissolved air in the water and finally become non-oscillatory (phase (iv)). The bubbles then grow through further evaporation and uptake of air molecules.
This description of the early stages of bubble formation is consistent with that reported by Wang et al.\cite{WangY2018}
The formation of giant bubbles in phase (ii)\cite{WangY2018,LiXiaolai2019,Detert2020} and the bubble growth primarily observed in phase (iv)\cite{Baffou2014,WangY2017} have been intensively studied by other research groups. 
However, the oscillating bubble growth in phase (iii) has not yet been fully investigated.
Bubble growth has been understood as the evaporation of water or the uptake of air molecules through diffusion and evaporation of water.
Herein, we discuss the exhalation and inhalation of air-rich bubbles as a new mode of oscillating bubble growth.

\begin{figure*}[tbp]
\centerline{\includegraphics[bb = 0 0 518.84 399.93, width=15cm]{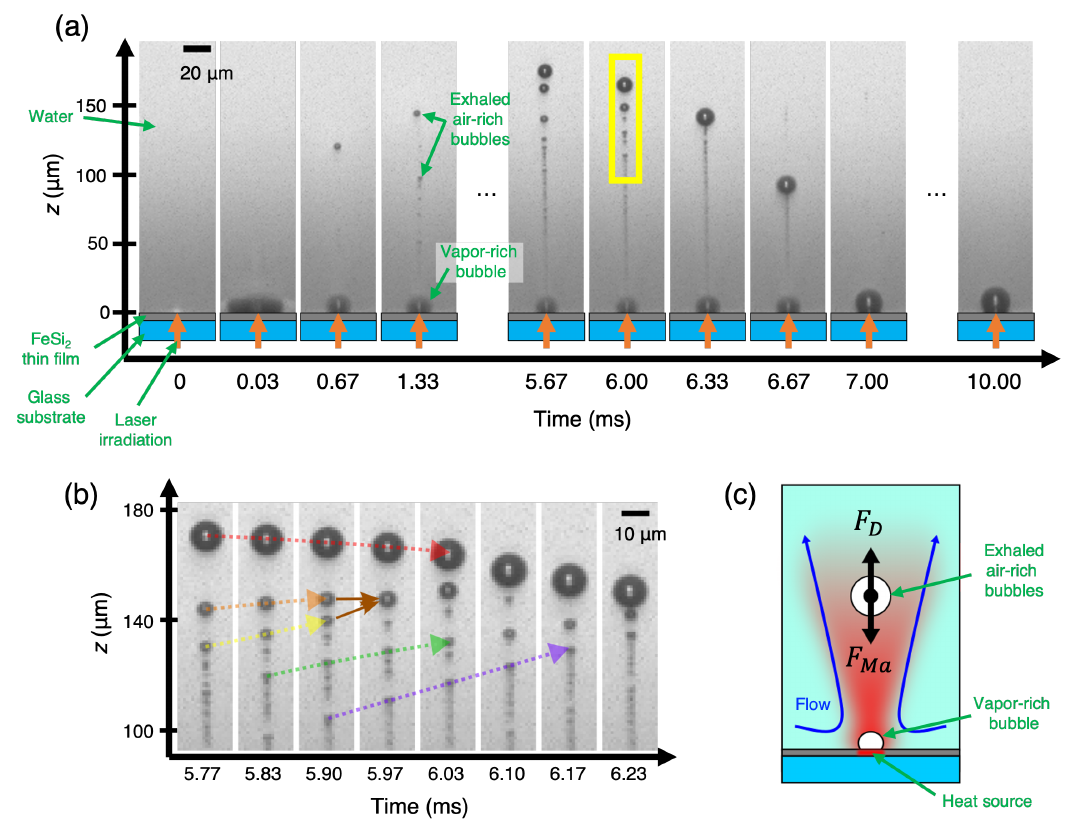}}
\caption{(a) Time-series image of bubble behavior from its initial formation, where $z$ is the distance from the substrate surface. (b) Detail of the bubble motion in the area marked by the yellow square in (a). (c) Schematic of the forces acting on the exhaled air-rich bubble.}
 \label{fig2}
\end{figure*}

Figure \ref{fig2}(a) shows a time-series image of the bubble formation and growth, focusing particularly on the oscillating bubbles (phase (iii)), 
where $z$ is the distance from the substrate surface.
We defined the time origin, i.e., $t = 0$, as the frame before the first bubble was observed after the start of photothermal heating.
After a certain waiting period following the start of photothermal heating, a water vapor-rich giant bubble was generated ($t = 0.03$ ms).
This corresponded to Phase (ii) in Fig. \ref{fig1}.
This bubble collapsed within $\sim$10 \si{\us}, leaving a self-oscillating water vapor-rich bubble of radius $\sim$9 \si{\um}.
Such oscillating water vapor-rich bubbles were characteristic of phase (iii) and were observed on the heated spot between $t$ = 0.67--6.67 ms.
Note that the frequency of the self-oscillation of the vapor-rich bubbles was on the order of sub-MHz and the period of oscillation was sufficiently shorter than the camera's exposure time.\cite{Li2017,Namura2020} Therefore, the bubbles do not appear to be oscillating in Fig. \ref{fig2}(a), but their outlines appear blurred.
As indicated in Fig \ref{fig2}(a), at $t$ = 1.33, the oscillating vapor-rich bubble exhaled tiny bubbles.
This bubble exhalation provided evidence that the bubbles were oscillating.
Please refer to the paper by Li et al. for a detailed report on an oscillating vapor-rich bubble exhaling small bubbles captured by an ultrahigh-speed camera.\cite{Li2017}
In this case, because bubbles were generated in non-degassed water, they continuously absorbed dissolved air from the water as the dissolved gas diffused and the water evaporated.
Vapor-rich bubbles exhale the absorbed air as air-rich bubbles because of their violent oscillations.
If the heat generation density is high, this cycle continues and the interior of the bubble remains vapor-rich for a long time.\cite{Li2017}
However, the laser power used in this study was too low to maintain this state for an extended period.
The exhaled air-rich bubbles fused and became bubbles with radii of approximately 7 \si{\um} (Fig. \ref{fig2}(a), $t$ = 5.67 ms).
The grown air-rich bubbles was then attracted to the vapor-rich bubbles at the heating point ($t$ = 5.67--6.67 ms).
The fusion of these bubbles caused the bubbles at the heating point to begin to grow in a manner characteristic of air-rich bubbles in phase (iv) ($t$ = 7.00--10.00 ms).

Figure \ref{fig2}(b) shows a more detailed temporal change during $t$ = 5.77--6.23 ms obtained by enlarging the yellow frame in Fig. \ref{fig2}(a). 
The image interval is 0.067 ms, and $z$ is the distance from the substrate surface. The dotted arrows indicate the change in the position of the same bubble over time. 
As indicated by the purple arrow, the relatively small bubbles moved away from the substrate surface. However, as indicated by the green, yellow, and orange arrows, as the bubbles grew, they moved away from the thin-film surface at a slower rate. Furthermore, a bubble with a radius of approximately 7 \si{\um} was attracted to the substrate.
This change in velocity with bubble size promoted the fusion of bubbles of different sizes. For example, the brown arrows in Fig. \ref{fig2}(b) indicate the moment of fusion of two bubbles.
Eventually, the largest bubble approached the substrate surface while swallowing other bubbles and fused with a vapor-rich bubbles on the substrate surface. The vapor-rich bubble was then fed with a large amount of air and expanded into an air-rich bubble.
This phenomenon is because the direction and strength of forces acting on the bubbles depend on their size. 
It is known that vapor-rich bubbles at the local heating point generate strong convection currents in the direction perpendicular to the substrate surface (Fig. \ref{fig2}(c)). Bubbles with diameters of approximately 1 \si{\um} are considered to be pushed away from the substrate by the flow created by the vapor-rich bubbles. However, as the bubbles grow, the influence of the other forces becomes pronounced, and air-rich bubbles are attracted to the substrate surface against the flow.

In the following section, we discuss why bubbles are attracted to the substrate surface.
Note that the observation plane is perpendicular to the direction of gravitational acceleration. Therefore, the bubble motion is not significantly affected by gravity. The main forces acting on the bubble are the drag force from the flow and the Marangoni force. First, the drag force $F_D$ depends on the relative velocity between the bubble and surrounding fluid. As shown in previous studies,\cite{Li2017,Namura2017} a vapor-rich bubble on the local heating point may generate a flow on the order of 1 mm/s to 1 m/s. Let $U_b$ be the speed at which the bubble moves, $U_0$ the flow speed of the surrounding water, and $U = U_0-U_b$ the relative velocity. The drag force is represented by $F_D = 1/2 \pi C_D \rho U^2 R^2 $, where $C_D$ is the drag coefficient, $\rho$ is the density of water, and $R$ is the bubble radius. Here, the direction of the $z$-axis is assumed as the positive direction (Fig. \ref{fig2}(c)). The drag coefficient is expressed as a function of the Reynolds number ($Re$), and its expression depends on the range of the $Re$ magnitude. This study employed $C_D=16/Re$, which is primarily used for $Re < 1$.\cite{Tomiyama1998} Because $Re= 2 \rho U R/\mu$, where $\mu$ is the dynamic viscosity of water, the drag force can be transformed as $F_D = 4 \pi \mu R U $. Therefore, the drag force is proportional to the relative velocity and bubble radius.

The Marangoni force $F_{Ma}$ is caused by the surface-tension gradient created by the temperature gradient. The heat source is a laser spot on the FeSi$_2$ thin film, and the temperature decreases the further away from the spot. Let the temperature gradient in the $z$ direction be $dT/dz$ and the temperature dependence of the surface tension be $d\sigma/dT$. For simplicity, we assume that the temperature gradient in the $z$ direction changes slowly and that $dT/dz$ on the bubble is constant. Subsequently, integrating the Marangoni force over the bubble surface yields $F_{Ma}= \frac{8}{3} \pi \frac{dT}{dz} \frac{d\sigma}{dT} R^2$. 
In this study, because $d\sigma/dT < 0$ and $dT/dz < 0$, $F_{Ma} > 0$. However, this force acts on the surrounding fluid, and the bubble moves in the opposite direction owing to the counteraction.  As shown in Fig. \ref{fig2}(c), the Marangoni force acting on the bubble is toward the heat source. Therefore, $F_{Ma}$ is positive in the negative direction along the $z$ axis. For water, $d\sigma/dT$ is almost constant regardless of temperature; thus, the Marangoni force is proportional to the temperature gradient and the square of the bubble radius. Because the drag force is proportional to the bubble radius, the Marangoni force tends to dominate the drag force as the bubble size increases.

Closely reexamining the bubble motion in Fig. \ref{fig2}(b), the bubble does not appear to be strongly accelerated in the $z$= 100--180 \si{\um} range. In other words, the Marangoni and drag forces are balanced at each position. Subsequently, $F_D = F_{Ma}$ can be rearranged as
\begin{equation}
  U = U_0 - U_b =  \frac{2}{3} \frac{R}{\mu} \frac{dT}{dz} \frac{d\sigma}{dT}.  \label{eq:1}
\end{equation}
In this equation, $U_0$ and $dT/dz$ can be regarded as constants if the value of $z$ is fixed.
Therefore, if we measure various combinations of $R$ and $U_b$ at a certain position $z$, we can fit Equation \ref{eq:1} to the measurement results using $U_0$ and $dT/dz$ as the fitting parameters.

\begin{figure}[tbp]
\centerline{\includegraphics[bb = 0 0 360 252, width=8cm]{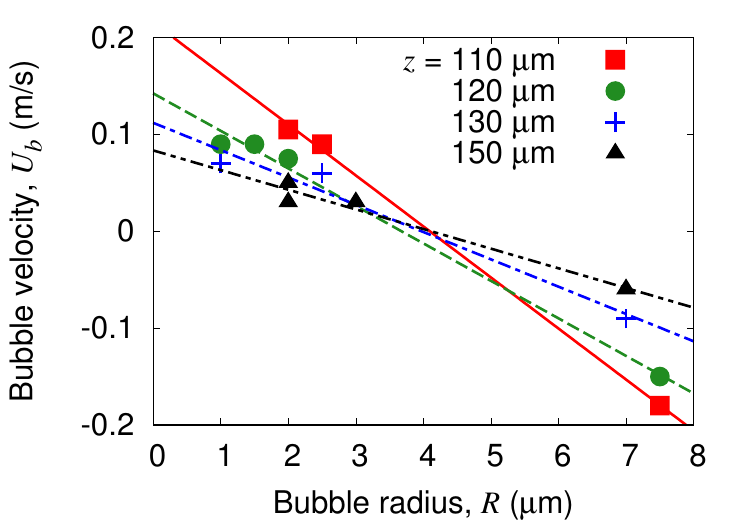}}
\caption{Red squares, green circles, blue crosses, and black triangles are the bubble radii and velocities measured at $z$ = 110, 120, 130, and 150 \si{\um}, respectively. The results of fitting Equation \ref{eq:1} to these results are indicated as lines.}
 \label{fig3}
\end{figure}

\begin{figure}[tbp]
\centerline{\includegraphics[bb = 0 0 345.6 491.52, width=7cm]{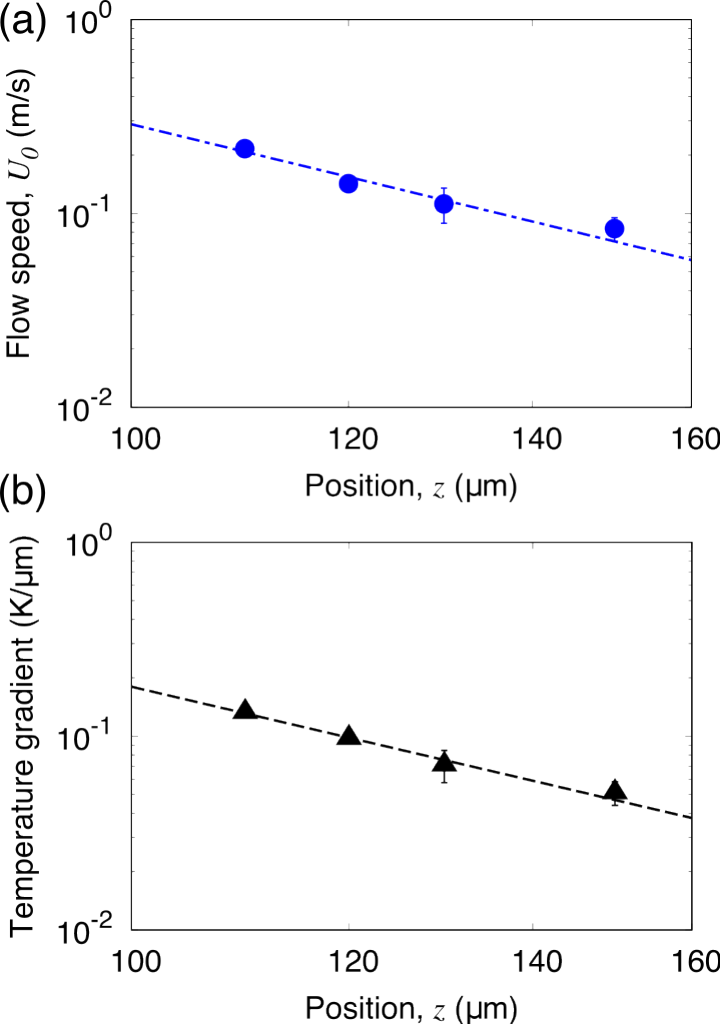}}
\caption{Fitting results of (a) flow speed $U_0$ and (b) temperature gradient $dT/dz$ as functions of the position $z$.}
 \label{fig4}
\end{figure}

The red square, green circle, blue cross, and black triangle in Fig. \ref{fig3} show the measured bubble radii and velocities at $z$ = 110, 120, 130, and 150 \si{\um}, respectively. 
The horizontal and vertical axes represent the bubble radius and speed of bubble motion, respectively. The lines show the results of fitting Equation \ref{eq:1} to these results, where $U_0$ and $dT/dz$ are the fitting parameters. Here, we employed $d\sigma/dT$ = $-$1.7$\times 10^{-4}$ N/m/K and $\mu$ = 2.8$\times 10^{-4}$ Pa s.
The solid red, dashed green, dash-dotted blue, and dash-dotted black lines correspond to the results for $z$ = 110, 120, 130, and 150 \si{\um}, respectively.
The smaller $z$, i.e., the closer it is to the heat source or vapor-rich bubble,
the larger the absolute value of the slope of the lines.
 The intercepts of the straight lines in the fitting results correspond to $U_0$, and the slopes are proportional to $dT/dz$. Figure \ref{fig4} depicts the results for the fitting parameters.
Figure \ref{fig4}(a) shows $U_0$ obtained through fitting as a function of $z$. The straight line represents the result of fitting a power function of $-3.4 \pm 0.54$. This result was consistent with the flow pattern around vapor-rich bubbles reported in our previous study,\cite{Namura2017} which showed that $U_0 \propto z^{-3}$.
Figure \ref{fig4}(b) shows $dT/dz$ obtained through fitting as a function of $z$. 
The temperature gradient increased as it approaches the heat source. This was reasonable as the heat source was small and the heat spread three-dimensionally.
The straight line represents the result of the power function fitting.  
Considering the temperature distribution in water in the presence of phase transitions and a strong flow created by vapor-rich bubbles is difficult. However, the temperature gradient was found to be proportional to $z$ to the $-3.3 \pm 0.27$th power.
These results suggest that Equation \ref{eq:1} agrees well with the experimental results.
In other words, the drag force and Marangoni force determine the motion of an air-rich bubble exhaled from a vapor-rich bubble. 
The reduction in Marangoni forces may result in the maintenance of vapor-rich bubbles that produce a strong flow.

\section{CONCLUSION}
In this study, the formation of air-rich bubbles at the local heating points was investigated. Immediately after bubble formation, an oscillating vapor-rich bubble was generated, and air-rich bubbles were exhaled from the vapor-rich bubble. 
The exhaled-air-rich bubbles fused with each other and gradually increased in size. When the radius reached 7 \si{\um}, the air-rich bubbles were attracted to the heat source against the flow created by the vapor-rich bubbles. As a result, these bubbles merged, and air-rich bubbles began to grow at the heat source.
The experimental bubble motion at $z$ = 110--150 \si{\um} can be well explained using the equation for the balance between the drag and Marangoni forces acting on air-rich bubbles.
Because the drag force is directly proportional to the bubble radius, and the Marangoni force is proportional to the square of the bubble radius, the larger the bubble, the stronger the influence of the Marangoni force.
Therefore, as the bubble grows, it is attracted to the heat source against the flow generated by the vapor-rich bubble. 
Additionally, the $z$-dependence of the velocity and temperature gradient at $z$ = 110--150 \si{\um} can be estimated to be proportional to $z$ to the minus 3.4--3.3 power. 
These results are useful for the better understanding of bubble growth and determining the conditions for the stable formation of vapor-rich bubbles in non-degassed liquids.

\section*{Acknowledgments}
This study was supported by JSPS KAKENHI Grant No. 21H01784 and the JST FOREST Program (Grant Number JPMJFR203N, Japan). It was also financially supported by a collaborative research project between Kyoto University and Mitsubishi Electric Corporation on Evolutionary Mechanical System Technology.


\bibliography{2023bubblegrowth}

\begin{thebibliography}{41}%
\makeatletter
\providecommand \@ifxundefined [1]{%
 \@ifx{#1\undefined}
}%
\providecommand \@ifnum [1]{%
 \ifnum #1\expandafter \@firstoftwo
 \else \expandafter \@secondoftwo
 \fi
}%
\providecommand \@ifx [1]{%
 \ifx #1\expandafter \@firstoftwo
 \else \expandafter \@secondoftwo
 \fi
}%
\providecommand \natexlab [1]{#1}%
\providecommand \enquote  [1]{``#1''}%
\providecommand \bibnamefont  [1]{#1}%
\providecommand \bibfnamefont [1]{#1}%
\providecommand \citenamefont [1]{#1}%
\providecommand \href@noop [0]{\@secondoftwo}%
\providecommand \href [0]{\begingroup \@sanitize@url \@href}%
\providecommand \@href[1]{\@@startlink{#1}\@@href}%
\providecommand \@@href[1]{\endgroup#1\@@endlink}%
\providecommand \@sanitize@url [0]{\catcode `\\12\catcode `\$12\catcode
  `\&12\catcode `\#12\catcode `\^12\catcode `\_12\catcode `\%12\relax}%
\providecommand \@@startlink[1]{}%
\providecommand \@@endlink[0]{}%
\providecommand \url  [0]{\begingroup\@sanitize@url \@url }%
\providecommand \@url [1]{\endgroup\@href {#1}{\urlprefix }}%
\providecommand \urlprefix  [0]{URL }%
\providecommand \Eprint [0]{\href }%
\providecommand \doibase [0]{http://dx.doi.org/}%
\providecommand \selectlanguage [0]{\@gobble}%
\providecommand \bibinfo  [0]{\@secondoftwo}%
\providecommand \bibfield  [0]{\@secondoftwo}%
\providecommand \translation [1]{[#1]}%
\providecommand \BibitemOpen [0]{}%
\providecommand \bibitemStop [0]{}%
\providecommand \bibitemNoStop [0]{.\EOS\space}%
\providecommand \EOS [0]{\spacefactor3000\relax}%
\providecommand \BibitemShut  [1]{\csname bibitem#1\endcsname}%
\let\auto@bib@innerbib\@empty
\bibitem [{\citenamefont {Geng}\ \emph {et~al.}(2001)\citenamefont {Geng},
  \citenamefont {Yuan}, \citenamefont {Oguz},\ and\ \citenamefont
  {Prosperetti}}]{Geng2001}%
  \BibitemOpen
  \bibfield  {author} {\bibinfo {author} {\bibfnamefont {X.}~\bibnamefont
  {Geng}}, \bibinfo {author} {\bibfnamefont {H.}~\bibnamefont {Yuan}}, \bibinfo
  {author} {\bibfnamefont {H.~N.}\ \bibnamefont {Oguz}}, \ and\ \bibinfo
  {author} {\bibfnamefont {A.}~\bibnamefont {Prosperetti}},\ }\bibfield
  {title} {\enquote {\bibinfo {title} {Bubble-based micropump for electrically
  conducting liquids},}\ }\href {http://stacks.iop.org/0960-1317/11/i=3/a=317}
  {\bibfield  {journal} {\bibinfo  {journal} {J. Micromech. Microeng.}\
  }\textbf {\bibinfo {volume} {11}},\ \bibinfo {pages} {270} (\bibinfo {year}
  {2001})}\BibitemShut {NoStop}%
\bibitem [{\citenamefont {Tho}, \citenamefont {Manasseh},\ and\ \citenamefont
  {Ooi}(2007)}]{Tho2007}%
  \BibitemOpen
  \bibfield  {author} {\bibinfo {author} {\bibfnamefont {P.}~\bibnamefont
  {Tho}}, \bibinfo {author} {\bibfnamefont {R.}~\bibnamefont {Manasseh}}, \
  and\ \bibinfo {author} {\bibfnamefont {A.}~\bibnamefont {Ooi}},\ }\bibfield
  {title} {\enquote {\bibinfo {title} {Cavitation microstreaming patterns in
  single and multiple bubble systems},}\ }\href {\doibase
  10.1017/S0022112006004393} {\bibfield  {journal} {\bibinfo  {journal} {J.
  Fluid Mech.}\ }\textbf {\bibinfo {volume} {576}},\ \bibinfo {pages}
  {191--233} (\bibinfo {year} {2007})}\BibitemShut {NoStop}%
\bibitem [{\citenamefont {Dijkink}\ and\ \citenamefont
  {Ohl}(2008)}]{Dijkink2008}%
  \BibitemOpen
  \bibfield  {author} {\bibinfo {author} {\bibfnamefont {R.}~\bibnamefont
  {Dijkink}}\ and\ \bibinfo {author} {\bibfnamefont {C.~D.}\ \bibnamefont
  {Ohl}},\ }\bibfield  {title} {\enquote {\bibinfo {title} {Laser-induced
  cavitation based micropump},}\ }\href {\doibase 10.1039/b806912c} {\bibfield
  {journal} {\bibinfo  {journal} {Lab Chip}\ }\textbf {\bibinfo {volume} {8}},\
  \bibinfo {pages} {1676--81} (\bibinfo {year} {2008})}\BibitemShut {NoStop}%
\bibitem [{\citenamefont {Patel}\ \emph {et~al.}(2014)\citenamefont {Patel},
  \citenamefont {Nanayakkara}, \citenamefont {Simon},\ and\ \citenamefont
  {Lee}}]{Patel2014}%
  \BibitemOpen
  \bibfield  {author} {\bibinfo {author} {\bibfnamefont {M.~V.}\ \bibnamefont
  {Patel}}, \bibinfo {author} {\bibfnamefont {I.~A.}\ \bibnamefont
  {Nanayakkara}}, \bibinfo {author} {\bibfnamefont {M.~G.}\ \bibnamefont
  {Simon}}, \ and\ \bibinfo {author} {\bibfnamefont {A.~P.}\ \bibnamefont
  {Lee}},\ }\bibfield  {title} {\enquote {\bibinfo {title} {Cavity-induced
  microstreaming for simultaneous on-chip pumping and size-based separation of
  cells and particles},}\ }\href {\doibase 10.1039/C4LC00447G} {\bibfield
  {journal} {\bibinfo  {journal} {Lab Chip}\ }\textbf {\bibinfo {volume}
  {14}},\ \bibinfo {pages} {3860--3872} (\bibinfo {year} {2014})}\BibitemShut
  {NoStop}%
\bibitem [{\citenamefont {Liu}\ \emph {et~al.}(2002)\citenamefont {Liu},
  \citenamefont {Yang}, \citenamefont {Pindera}, \citenamefont {Athavale},\
  and\ \citenamefont {Grodzinski}}]{Liu2002}%
  \BibitemOpen
  \bibfield  {author} {\bibinfo {author} {\bibfnamefont {R.~H.}\ \bibnamefont
  {Liu}}, \bibinfo {author} {\bibfnamefont {J.}~\bibnamefont {Yang}}, \bibinfo
  {author} {\bibfnamefont {M.~Z.}\ \bibnamefont {Pindera}}, \bibinfo {author}
  {\bibfnamefont {M.}~\bibnamefont {Athavale}}, \ and\ \bibinfo {author}
  {\bibfnamefont {P.}~\bibnamefont {Grodzinski}},\ }\bibfield  {title}
  {\enquote {\bibinfo {title} {Bubble-induced acoustic micromixing},}\ }\href
  {\doibase 10.1039/B201952C} {\bibfield  {journal} {\bibinfo  {journal} {Lab
  Chip}\ }\textbf {\bibinfo {volume} {2}},\ \bibinfo {pages} {151--157}
  (\bibinfo {year} {2002})}\BibitemShut {NoStop}%
\bibitem [{\citenamefont {Hellman}\ \emph {et~al.}(2007)\citenamefont
  {Hellman}, \citenamefont {Rau}, \citenamefont {Yoon}, \citenamefont {Bae},
  \citenamefont {Palmer}, \citenamefont {Phillips}, \citenamefont
  {Allbritton},\ and\ \citenamefont {Venugopalan}}]{Hellman2007}%
  \BibitemOpen
  \bibfield  {author} {\bibinfo {author} {\bibfnamefont {A.~N.}\ \bibnamefont
  {Hellman}}, \bibinfo {author} {\bibfnamefont {K.~R.}\ \bibnamefont {Rau}},
  \bibinfo {author} {\bibfnamefont {H.~H.}\ \bibnamefont {Yoon}}, \bibinfo
  {author} {\bibfnamefont {S.}~\bibnamefont {Bae}}, \bibinfo {author}
  {\bibfnamefont {J.~F.}\ \bibnamefont {Palmer}}, \bibinfo {author}
  {\bibfnamefont {K.~S.}\ \bibnamefont {Phillips}}, \bibinfo {author}
  {\bibfnamefont {N.~L.}\ \bibnamefont {Allbritton}}, \ and\ \bibinfo {author}
  {\bibfnamefont {V.}~\bibnamefont {Venugopalan}},\ }\bibfield  {title}
  {\enquote {\bibinfo {title} {Laser-induced mixing in microfluidic
  channels},}\ }\href {\doibase 10.1021/ac070081i} {\bibfield  {journal}
  {\bibinfo  {journal} {Anal. Chem.}\ }\textbf {\bibinfo {volume} {79}},\
  \bibinfo {pages} {4484--4492} (\bibinfo {year} {2007})}\BibitemShut {NoStop}%
\bibitem [{\citenamefont {Serra}\ and\ \citenamefont
  {Piqu^^c3^^a9}(2019)}]{Serra2019}%
  \BibitemOpen
  \bibfield  {author} {\bibinfo {author} {\bibfnamefont {P.}~\bibnamefont
  {Serra}}\ and\ \bibinfo {author} {\bibfnamefont {A.}~\bibnamefont
  {Piqu^^c3^^a9}},\ }\bibfield  {title} {\enquote {\bibinfo {title}
  {Laser-induced forward transfer: Fundamentals and applications},}\ }\href
  {\doibase https://doi.org/10.1002/admt.201800099} {\bibfield  {journal}
  {\bibinfo  {journal} {Adv. Mater. Technol.}\ }\textbf {\bibinfo {volume}
  {4}},\ \bibinfo {pages} {1800099} (\bibinfo {year} {2019})}\BibitemShut
  {NoStop}%
\bibitem [{\citenamefont {Kim}\ \emph {et~al.}(2009)\citenamefont {Kim},
  \citenamefont {Galanzha}, \citenamefont {Shashkov}, \citenamefont {Moon},\
  and\ \citenamefont {Zharov}}]{Kim2009}%
  \BibitemOpen
  \bibfield  {author} {\bibinfo {author} {\bibfnamefont {J.-W.}\ \bibnamefont
  {Kim}}, \bibinfo {author} {\bibfnamefont {E.~I.}\ \bibnamefont {Galanzha}},
  \bibinfo {author} {\bibfnamefont {E.~V.}\ \bibnamefont {Shashkov}}, \bibinfo
  {author} {\bibfnamefont {H.-M.}\ \bibnamefont {Moon}}, \ and\ \bibinfo
  {author} {\bibfnamefont {V.~P.}\ \bibnamefont {Zharov}},\ }\bibfield  {title}
  {\enquote {\bibinfo {title} {Golden carbon nanotubes as multimodal
  photoacoustic and photothermal high-contrast molecular agents},}\ }\href
  {\doibase
  http://www.nature.com/nnano/journal/v4/n10/suppinfo/nnano.2009.231_S1.html}
  {\bibfield  {journal} {\bibinfo  {journal} {Nat. Nano}\ }\textbf {\bibinfo
  {volume} {4}},\ \bibinfo {pages} {688--694} (\bibinfo {year}
  {2009})}\BibitemShut {NoStop}%
\bibitem [{\citenamefont {Ju}, \citenamefont {Roy},\ and\ \citenamefont
  {Murray}(2013)}]{Ju2013}%
  \BibitemOpen
  \bibfield  {author} {\bibinfo {author} {\bibfnamefont {H.}~\bibnamefont
  {Ju}}, \bibinfo {author} {\bibfnamefont {R.~A.}\ \bibnamefont {Roy}}, \ and\
  \bibinfo {author} {\bibfnamefont {T.~W.}\ \bibnamefont {Murray}},\ }\bibfield
   {title} {\enquote {\bibinfo {title} {Gold nanoparticle targeted
  photoacoustic cavitation for potential deep tissue imaging and therapy},}\
  }\href {\doibase 10.1364/BOE.4.000066} {\bibfield  {journal} {\bibinfo
  {journal} {Biomed. Opt. Express}\ }\textbf {\bibinfo {volume} {4}},\ \bibinfo
  {pages} {66--76} (\bibinfo {year} {2013})}\BibitemShut {NoStop}%
\bibitem [{\citenamefont {Ortega-Mendoza}\ \emph {et~al.}(2018)\citenamefont
  {Ortega-Mendoza}, \citenamefont {Sarabia-Alonso}, \citenamefont
  {Zaca-Mor^^c3^^a1n}, \citenamefont {Padilla-Vivanco}, \citenamefont
  {Toxqui-Quitl}, \citenamefont {Rivas-Cambero}, \citenamefont
  {Ramirez-Ramirez}, \citenamefont {Torres-Hurtado},\ and\ \citenamefont
  {Ramos-Garc^^c3^^ada}}]{Ortega2018}%
  \BibitemOpen
  \bibfield  {author} {\bibinfo {author} {\bibfnamefont {J.~G.}\ \bibnamefont
  {Ortega-Mendoza}}, \bibinfo {author} {\bibfnamefont {J.~A.}\ \bibnamefont
  {Sarabia-Alonso}}, \bibinfo {author} {\bibfnamefont {P.}~\bibnamefont
  {Zaca-Mor^^c3^^a1n}}, \bibinfo {author} {\bibfnamefont {A.}~\bibnamefont
  {Padilla-Vivanco}}, \bibinfo {author} {\bibfnamefont {C.}~\bibnamefont
  {Toxqui-Quitl}}, \bibinfo {author} {\bibfnamefont {I.}~\bibnamefont
  {Rivas-Cambero}}, \bibinfo {author} {\bibfnamefont {J.}~\bibnamefont
  {Ramirez-Ramirez}}, \bibinfo {author} {\bibfnamefont {S.~A.}\ \bibnamefont
  {Torres-Hurtado}}, \ and\ \bibinfo {author} {\bibfnamefont {R.}~\bibnamefont
  {Ramos-Garc^^c3^^ada}},\ }\bibfield  {title} {\enquote {\bibinfo {title}
  {Marangoni force-driven manipulation of photothermally-induced
  microbubbles},}\ }\href {\doibase 10.1364/OE.26.006653} {\bibfield  {journal}
  {\bibinfo  {journal} {Opt. Express}\ }\textbf {\bibinfo {volume} {26}},\
  \bibinfo {pages} {6653--6662} (\bibinfo {year} {2018})}\BibitemShut {NoStop}%
\bibitem [{\citenamefont {Sarabia-Alonso}\ \emph {et~al.}(2021)\citenamefont
  {Sarabia-Alonso}, \citenamefont {Ortega-Mendoza}, \citenamefont {Mansurova},
  \citenamefont {Mu^^c3^^b1oz-P^^c3^^a9rez},\ and\ \citenamefont
  {Ramos-Garc^^c3^^ada}}]{Sarabia2021}%
  \BibitemOpen
  \bibfield  {author} {\bibinfo {author} {\bibfnamefont {J.~A.}\ \bibnamefont
  {Sarabia-Alonso}}, \bibinfo {author} {\bibfnamefont {J.~G.}\ \bibnamefont
  {Ortega-Mendoza}}, \bibinfo {author} {\bibfnamefont {S.}~\bibnamefont
  {Mansurova}}, \bibinfo {author} {\bibfnamefont {F.~M.}\ \bibnamefont
  {Mu^^c3^^b1oz-P^^c3^^a9rez}}, \ and\ \bibinfo {author} {\bibfnamefont
  {R.}~\bibnamefont {Ramos-Garc^^c3^^ada}},\ }\bibfield  {title} {\enquote
  {\bibinfo {title} {3d trapping of microbubbles by the marangoni force},}\
  }\href {\doibase 10.1364/OL.440290} {\bibfield  {journal} {\bibinfo
  {journal} {Opt. Lett.}\ }\textbf {\bibinfo {volume} {46}},\ \bibinfo {pages}
  {5786--5789} (\bibinfo {year} {2021})}\BibitemShut {NoStop}%
\bibitem [{\citenamefont {Berry}, \citenamefont {Heckenberg},\ and\
  \citenamefont {Rubinszteindunlop}(2000)}]{Berry2000}%
  \BibitemOpen
  \bibfield  {author} {\bibinfo {author} {\bibfnamefont {D.~W.}\ \bibnamefont
  {Berry}}, \bibinfo {author} {\bibfnamefont {N.~R.}\ \bibnamefont
  {Heckenberg}}, \ and\ \bibinfo {author} {\bibfnamefont {H.}~\bibnamefont
  {Rubinszteindunlop}},\ }\bibfield  {title} {\enquote {\bibinfo {title}
  {Effects associated with bubble formation in optical trapping},}\ }\href
  {\doibase 10.1080/09500340008235124} {\bibfield  {journal} {\bibinfo
  {journal} {J. Mod. Opt.}\ }\textbf {\bibinfo {volume} {47}},\ \bibinfo
  {pages} {1575--1585} (\bibinfo {year} {2000})}\BibitemShut {NoStop}%
\bibitem [{\citenamefont {Namura}\ \emph {et~al.}(2015)\citenamefont {Namura},
  \citenamefont {Nakajima}, \citenamefont {Kimura},\ and\ \citenamefont
  {Suzuki}}]{Namura2015}%
  \BibitemOpen
  \bibfield  {author} {\bibinfo {author} {\bibfnamefont {K.}~\bibnamefont
  {Namura}}, \bibinfo {author} {\bibfnamefont {K.}~\bibnamefont {Nakajima}},
  \bibinfo {author} {\bibfnamefont {K.}~\bibnamefont {Kimura}}, \ and\ \bibinfo
  {author} {\bibfnamefont {M.}~\bibnamefont {Suzuki}},\ }\bibfield  {title}
  {\enquote {\bibinfo {title} {Photothermally controlled {M}arangoni flow
  around a micro bubble},}\ }\href@noop {} {\bibfield  {journal} {\bibinfo
  {journal} {Appl. Phys. Lett.}\ }\textbf {\bibinfo {volume} {106}},\ \bibinfo
  {pages} {043101} (\bibinfo {year} {2015})}\BibitemShut {NoStop}%
\bibitem [{\citenamefont {Lee}\ and\ \citenamefont {Luo}(2020)}]{Lee2020}%
  \BibitemOpen
  \bibfield  {author} {\bibinfo {author} {\bibfnamefont {E.}~\bibnamefont
  {Lee}}\ and\ \bibinfo {author} {\bibfnamefont {T.}~\bibnamefont {Luo}},\
  }\bibfield  {title} {\enquote {\bibinfo {title} {Long-distance optical
  pulling of nanoparticle in a low index cavity using a single plane wave},}\
  }\href {\doibase doi:10.1126/sciadv.aaz3646} {\bibfield  {journal} {\bibinfo
  {journal} {Sci. Adv.}\ }\textbf {\bibinfo {volume} {6}},\ \bibinfo {pages}
  {eaaz3646} (\bibinfo {year} {2020})}\BibitemShut {NoStop}%
\bibitem [{\citenamefont {Dara}\ \emph {et~al.}(2023)\citenamefont {Dara},
  \citenamefont {Shanei}, \citenamefont {Jones},\ and\ \citenamefont
  {K\"{a}ll}}]{Dara2023}%
  \BibitemOpen
  \bibfield  {author} {\bibinfo {author} {\bibfnamefont {P.}~\bibnamefont
  {Dara}}, \bibinfo {author} {\bibfnamefont {M.}~\bibnamefont {Shanei}},
  \bibinfo {author} {\bibfnamefont {S.}~\bibnamefont {Jones}}, \ and\ \bibinfo
  {author} {\bibfnamefont {M.}~\bibnamefont {K\"{a}ll}},\ }\bibfield  {title}
  {\enquote {\bibinfo {title} {Directional control of transient flows generated
  by thermoplasmonic bubble nucleation},}\ }\href {\doibase
  10.1021/acs.jpcc.3c02263} {\bibfield  {journal} {\bibinfo  {journal} {J.
  Phys. Chem. C}\ }\textbf {\bibinfo {volume} {127}},\ \bibinfo {pages}
  {17454--17459} (\bibinfo {year} {2023})}\BibitemShut {NoStop}%
\bibitem [{\citenamefont {Taylor}\ and\ \citenamefont
  {Hnatovsky}(2004)}]{Taylor2004}%
  \BibitemOpen
  \bibfield  {author} {\bibinfo {author} {\bibfnamefont {R.}~\bibnamefont
  {Taylor}}\ and\ \bibinfo {author} {\bibfnamefont {C.}~\bibnamefont
  {Hnatovsky}},\ }\bibfield  {title} {\enquote {\bibinfo {title} {Trapping and
  mixing of particles in water using a microbubble attached to an nsom fiber
  probe},}\ }\href {\doibase 10.1364/OPEX.12.000916} {\bibfield  {journal}
  {\bibinfo  {journal} {Opt. Express}\ }\textbf {\bibinfo {volume} {12}},\
  \bibinfo {pages} {916--928} (\bibinfo {year} {2004})}\BibitemShut {NoStop}%
\bibitem [{\citenamefont {Jones}\ \emph {et~al.}(2020)\citenamefont {Jones},
  \citenamefont {Andr\'{e}n}, \citenamefont {Antosiewicz}, \citenamefont
  {Stilgoe}, \citenamefont {Rubinsztein-Dunlop},\ and\ \citenamefont
  {K\"{a}ll}}]{Jones2020}%
  \BibitemOpen
  \bibfield  {author} {\bibinfo {author} {\bibfnamefont {S.}~\bibnamefont
  {Jones}}, \bibinfo {author} {\bibfnamefont {D.}~\bibnamefont {Andr\'{e}n}},
  \bibinfo {author} {\bibfnamefont {T.~J.}\ \bibnamefont {Antosiewicz}},
  \bibinfo {author} {\bibfnamefont {A.}~\bibnamefont {Stilgoe}}, \bibinfo
  {author} {\bibfnamefont {H.}~\bibnamefont {Rubinsztein-Dunlop}}, \ and\
  \bibinfo {author} {\bibfnamefont {M.}~\bibnamefont {K\"{a}ll}},\ }\bibfield
  {title} {\enquote {\bibinfo {title} {Strong transient flows generated by
  thermoplasmonic bubble nucleation},}\ }\href
  {https://www.ncbi.nlm.nih.gov/pubmed/33290656} {\bibfield  {journal}
  {\bibinfo  {journal} {ACS Nano}\ }\textbf {\bibinfo {volume} {14}},\ \bibinfo
  {pages} {17468--17475} (\bibinfo {year} {2020})}\BibitemShut {NoStop}%
\bibitem [{\citenamefont {Lin}\ \emph {et~al.}(2016)\citenamefont {Lin},
  \citenamefont {Peng}, \citenamefont {Mao}, \citenamefont {Li}, \citenamefont
  {Yogeesh}, \citenamefont {Rajeeva}, \citenamefont {Perillo}, \citenamefont
  {Dunn}, \citenamefont {Akinwande},\ and\ \citenamefont
  {Zheng}}]{LinLinhan2016}%
  \BibitemOpen
  \bibfield  {author} {\bibinfo {author} {\bibfnamefont {L.}~\bibnamefont
  {Lin}}, \bibinfo {author} {\bibfnamefont {X.}~\bibnamefont {Peng}}, \bibinfo
  {author} {\bibfnamefont {Z.}~\bibnamefont {Mao}}, \bibinfo {author}
  {\bibfnamefont {W.}~\bibnamefont {Li}}, \bibinfo {author} {\bibfnamefont
  {M.~N.}\ \bibnamefont {Yogeesh}}, \bibinfo {author} {\bibfnamefont {B.~B.}\
  \bibnamefont {Rajeeva}}, \bibinfo {author} {\bibfnamefont {E.~P.}\
  \bibnamefont {Perillo}}, \bibinfo {author} {\bibfnamefont {A.~K.}\
  \bibnamefont {Dunn}}, \bibinfo {author} {\bibfnamefont {D.}~\bibnamefont
  {Akinwande}}, \ and\ \bibinfo {author} {\bibfnamefont {Y.}~\bibnamefont
  {Zheng}},\ }\bibfield  {title} {\enquote {\bibinfo {title} {Bubble-pen
  lithography},}\ }\href {\doibase 10.1021/acs.nanolett.5b04524} {\bibfield
  {journal} {\bibinfo  {journal} {Nano Lett.}\ }\textbf {\bibinfo {volume}
  {16}},\ \bibinfo {pages} {701--708} (\bibinfo {year} {2016})}\BibitemShut
  {NoStop}%
\bibitem [{\citenamefont {Fujii}\ \emph {et~al.}(2017)\citenamefont {Fujii},
  \citenamefont {Fukano}, \citenamefont {Hayami}, \citenamefont {Ozawa},
  \citenamefont {Muneyuki}, \citenamefont {Kitamura},\ and\ \citenamefont
  {Haga}}]{Fujii2017}%
  \BibitemOpen
  \bibfield  {author} {\bibinfo {author} {\bibfnamefont {S.}~\bibnamefont
  {Fujii}}, \bibinfo {author} {\bibfnamefont {R.}~\bibnamefont {Fukano}},
  \bibinfo {author} {\bibfnamefont {Y.}~\bibnamefont {Hayami}}, \bibinfo
  {author} {\bibfnamefont {H.}~\bibnamefont {Ozawa}}, \bibinfo {author}
  {\bibfnamefont {E.}~\bibnamefont {Muneyuki}}, \bibinfo {author}
  {\bibfnamefont {N.}~\bibnamefont {Kitamura}}, \ and\ \bibinfo {author}
  {\bibfnamefont {M.-a.}\ \bibnamefont {Haga}},\ }\bibfield  {title} {\enquote
  {\bibinfo {title} {Simultaneous formation and spatial patterning of zno on
  ito surfaces by local laser-induced generation of microbubbles in aqueous
  solutions of [zn(nh3)4]2+},}\ }\href {\doibase 10.1021/acsami.6b16719}
  {\bibfield  {journal} {\bibinfo  {journal} {ACS Appl. Mater. Interfaces}\
  }\textbf {\bibinfo {volume} {9}},\ \bibinfo {pages} {8413--8419} (\bibinfo
  {year} {2017})}\BibitemShut {NoStop}%
\bibitem [{\citenamefont {Moon}\ \emph {et~al.}(2020)\citenamefont {Moon},
  \citenamefont {Zhang}, \citenamefont {Huang}, \citenamefont {Senapati},
  \citenamefont {Chang}, \citenamefont {Lee},\ and\ \citenamefont
  {Luo}}]{Moon2020}%
  \BibitemOpen
  \bibfield  {author} {\bibinfo {author} {\bibfnamefont {S.}~\bibnamefont
  {Moon}}, \bibinfo {author} {\bibfnamefont {Q.}~\bibnamefont {Zhang}},
  \bibinfo {author} {\bibfnamefont {D.}~\bibnamefont {Huang}}, \bibinfo
  {author} {\bibfnamefont {S.}~\bibnamefont {Senapati}}, \bibinfo {author}
  {\bibfnamefont {H.}~\bibnamefont {Chang}}, \bibinfo {author} {\bibfnamefont
  {E.}~\bibnamefont {Lee}}, \ and\ \bibinfo {author} {\bibfnamefont
  {T.}~\bibnamefont {Luo}},\ }\bibfield  {title} {\enquote {\bibinfo {title}
  {Biocompatible direct deposition of functionalized nanoparticles using
  shrinking surface plasmonic bubble},}\ }\href@noop {} {\bibfield  {journal}
  {\bibinfo  {journal} {Adv. Mater. Interfaces}\ }\textbf {\bibinfo {volume}
  {7}} (\bibinfo {year} {2020})}\BibitemShut {NoStop}%
\bibitem [{\citenamefont {Nishimura}\ \emph {et~al.}(2014)\citenamefont
  {Nishimura}, \citenamefont {Nishida}, \citenamefont {Yamamoto}, \citenamefont
  {Ito}, \citenamefont {Tokonami},\ and\ \citenamefont {Iida}}]{Nishimura2014}%
  \BibitemOpen
  \bibfield  {author} {\bibinfo {author} {\bibfnamefont {Y.}~\bibnamefont
  {Nishimura}}, \bibinfo {author} {\bibfnamefont {K.}~\bibnamefont {Nishida}},
  \bibinfo {author} {\bibfnamefont {Y.}~\bibnamefont {Yamamoto}}, \bibinfo
  {author} {\bibfnamefont {S.}~\bibnamefont {Ito}}, \bibinfo {author}
  {\bibfnamefont {S.}~\bibnamefont {Tokonami}}, \ and\ \bibinfo {author}
  {\bibfnamefont {T.}~\bibnamefont {Iida}},\ }\bibfield  {title} {\enquote
  {\bibinfo {title} {Control of submillimeter phase transition by collective
  photothermal effect},}\ }\href {\doibase 10.1021/jp506405w} {\bibfield
  {journal} {\bibinfo  {journal} {J. Phys. Chem. C}\ }\textbf {\bibinfo
  {volume} {118}},\ \bibinfo {pages} {18799--18804} (\bibinfo {year}
  {2014})}\BibitemShut {NoStop}%
\bibitem [{\citenamefont {Tokonami}\ \emph {et~al.}(2020)\citenamefont
  {Tokonami}, \citenamefont {Kurita}, \citenamefont {Yoshikawa}, \citenamefont
  {Sakurai}, \citenamefont {Suehiro}, \citenamefont {Yamamoto}, \citenamefont
  {Tamura}, \citenamefont {Karthaus},\ and\ \citenamefont
  {Iida}}]{Tokonami2020}%
  \BibitemOpen
  \bibfield  {author} {\bibinfo {author} {\bibfnamefont {S.}~\bibnamefont
  {Tokonami}}, \bibinfo {author} {\bibfnamefont {S.}~\bibnamefont {Kurita}},
  \bibinfo {author} {\bibfnamefont {R.}~\bibnamefont {Yoshikawa}}, \bibinfo
  {author} {\bibfnamefont {K.}~\bibnamefont {Sakurai}}, \bibinfo {author}
  {\bibfnamefont {T.}~\bibnamefont {Suehiro}}, \bibinfo {author} {\bibfnamefont
  {Y.}~\bibnamefont {Yamamoto}}, \bibinfo {author} {\bibfnamefont
  {M.}~\bibnamefont {Tamura}}, \bibinfo {author} {\bibfnamefont
  {O.}~\bibnamefont {Karthaus}}, \ and\ \bibinfo {author} {\bibfnamefont
  {T.}~\bibnamefont {Iida}},\ }\bibfield  {title} {\enquote {\bibinfo {title}
  {Light-induced assembly of living bacteria with honeycomb substrate},}\
  }\href {\doibase doi:10.1126/sciadv.aaz5757} {\bibfield  {journal} {\bibinfo
  {journal} {Sci. Adv.}\ }\textbf {\bibinfo {volume} {6}},\ \bibinfo {pages}
  {eaaz5757} (\bibinfo {year} {2020})}\BibitemShut {NoStop}%
\bibitem [{\citenamefont {Hou}\ \emph {et~al.}(2015)\citenamefont {Hou},
  \citenamefont {Yorulmaz}, \citenamefont {Verhart},\ and\ \citenamefont
  {Orrit}}]{Hou2015}%
  \BibitemOpen
  \bibfield  {author} {\bibinfo {author} {\bibfnamefont {L.}~\bibnamefont
  {Hou}}, \bibinfo {author} {\bibfnamefont {M.}~\bibnamefont {Yorulmaz}},
  \bibinfo {author} {\bibfnamefont {N.~R.}\ \bibnamefont {Verhart}}, \ and\
  \bibinfo {author} {\bibfnamefont {M.}~\bibnamefont {Orrit}},\ }\bibfield
  {title} {\enquote {\bibinfo {title} {Explosive formation and dynamics of
  vapor nanobubbles around a continuously heated gold nanosphere},}\ }\href
  {\doibase 10.1088/1367-2630/17/1/013050} {\bibfield  {journal} {\bibinfo
  {journal} {New J. Phys.}\ }\textbf {\bibinfo {volume} {17}},\ \bibinfo
  {pages} {013050} (\bibinfo {year} {2015})}\BibitemShut {NoStop}%
\bibitem [{\citenamefont {Jollans}\ and\ \citenamefont
  {Orrit}(2019)}]{Jollans2019}%
  \BibitemOpen
  \bibfield  {author} {\bibinfo {author} {\bibfnamefont {T.}~\bibnamefont
  {Jollans}}\ and\ \bibinfo {author} {\bibfnamefont {M.}~\bibnamefont
  {Orrit}},\ }\bibfield  {title} {\enquote {\bibinfo {title} {Explosive,
  oscillatory, and leidenfrost boiling at the nanoscale},}\ }\href {\doibase
  10.1103/PhysRevE.99.063110} {\bibfield  {journal} {\bibinfo  {journal} {Phys.
  Rev. E}\ }\textbf {\bibinfo {volume} {99}},\ \bibinfo {pages} {063110}
  (\bibinfo {year} {2019})}\BibitemShut {NoStop}%
\bibitem [{\citenamefont {Zeng}\ \emph {et~al.}(2021)\citenamefont {Zeng},
  \citenamefont {Chong}, \citenamefont {Wang}, \citenamefont {Diddens},
  \citenamefont {Li}, \citenamefont {Detert}, \citenamefont {Zandvliet},\ and\
  \citenamefont {Lohse}}]{Zeng2021}%
  \BibitemOpen
  \bibfield  {author} {\bibinfo {author} {\bibfnamefont {B.}~\bibnamefont
  {Zeng}}, \bibinfo {author} {\bibfnamefont {K.~L.}\ \bibnamefont {Chong}},
  \bibinfo {author} {\bibfnamefont {Y.}~\bibnamefont {Wang}}, \bibinfo {author}
  {\bibfnamefont {C.}~\bibnamefont {Diddens}}, \bibinfo {author} {\bibfnamefont
  {X.}~\bibnamefont {Li}}, \bibinfo {author} {\bibfnamefont {M.}~\bibnamefont
  {Detert}}, \bibinfo {author} {\bibfnamefont {H.~J.~W.}\ \bibnamefont
  {Zandvliet}}, \ and\ \bibinfo {author} {\bibfnamefont {D.}~\bibnamefont
  {Lohse}},\ }\bibfield  {title} {\enquote {\bibinfo {title} {Periodic bouncing
  of a plasmonic bubble in a binary liquid by competing solutal and thermal
  marangoni forces},}\ }\href {\doibase 10.1073/pnas.2103215118} {\bibfield
  {journal} {\bibinfo  {journal} {PNAS}\ }\textbf {\bibinfo {volume} {118}}
  (\bibinfo {year} {2021}),\ 10.1073/pnas.2103215118}\BibitemShut {NoStop}%
\bibitem [{\citenamefont {Christopher}, \citenamefont {Wang},\ and\
  \citenamefont {Peng}(2005)}]{Christopher2005}%
  \BibitemOpen
  \bibfield  {author} {\bibinfo {author} {\bibfnamefont {D.~M.}\ \bibnamefont
  {Christopher}}, \bibinfo {author} {\bibfnamefont {H.}~\bibnamefont {Wang}}, \
  and\ \bibinfo {author} {\bibfnamefont {X.~F.}\ \bibnamefont {Peng}},\
  }\bibfield  {title} {\enquote {\bibinfo {title} {Dynamics of bubble motion
  and bubble top jet flows from moving vapor bubbles on microwires},}\ }\href
  {\doibase 10.1115/1.2039109} {\bibfield  {journal} {\bibinfo  {journal} {J.
  Heat Trans.-T. ASME}\ }\textbf {\bibinfo {volume} {127}},\ \bibinfo {pages}
  {1260--1268} (\bibinfo {year} {2005})}\BibitemShut {NoStop}%
\bibitem [{\citenamefont {Li}\ \emph {et~al.}(2017)\citenamefont {Li},
  \citenamefont {Gonzalez-Avila}, \citenamefont {Nguyen},\ and\ \citenamefont
  {Ohl}}]{Li2017}%
  \BibitemOpen
  \bibfield  {author} {\bibinfo {author} {\bibfnamefont {F.}~\bibnamefont
  {Li}}, \bibinfo {author} {\bibfnamefont {S.~R.}\ \bibnamefont
  {Gonzalez-Avila}}, \bibinfo {author} {\bibfnamefont {D.~M.}\ \bibnamefont
  {Nguyen}}, \ and\ \bibinfo {author} {\bibfnamefont {C.-D.}\ \bibnamefont
  {Ohl}},\ }\bibfield  {title} {\enquote {\bibinfo {title} {Oscillate boiling
  from microheaters},}\ }\href {\doibase 10.1103/PhysRevFluids.2.014007}
  {\bibfield  {journal} {\bibinfo  {journal} {Phys. Rev. Fluid}\ }\textbf
  {\bibinfo {volume} {2}},\ \bibinfo {pages} {014007} (\bibinfo {year}
  {2017})}\BibitemShut {NoStop}%
\bibitem [{\citenamefont {Namura}, \citenamefont {Nakajima},\ and\
  \citenamefont {Suzuki}(2017)}]{Namura2017}%
  \BibitemOpen
  \bibfield  {author} {\bibinfo {author} {\bibfnamefont {K.}~\bibnamefont
  {Namura}}, \bibinfo {author} {\bibfnamefont {K.}~\bibnamefont {Nakajima}}, \
  and\ \bibinfo {author} {\bibfnamefont {M.}~\bibnamefont {Suzuki}},\
  }\bibfield  {title} {\enquote {\bibinfo {title} {Quasi-stokeslet induced by
  thermoplasmonic {M}arangoni effect around a water vapor microbubble},}\
  }\href {\doibase 10.1038/srep45776} {\bibfield  {journal} {\bibinfo
  {journal} {Sci. Rep.}\ }\textbf {\bibinfo {volume} {7}},\ \bibinfo {pages}
  {45776} (\bibinfo {year} {2017})}\BibitemShut {NoStop}%
\bibitem [{\citenamefont {Namura}\ \emph {et~al.}(2019)\citenamefont {Namura},
  \citenamefont {Imafuku}, \citenamefont {Kumar}, \citenamefont {Nakajima},
  \citenamefont {Sakakura},\ and\ \citenamefont {Suzuki}}]{Namura2019}%
  \BibitemOpen
  \bibfield  {author} {\bibinfo {author} {\bibfnamefont {K.}~\bibnamefont
  {Namura}}, \bibinfo {author} {\bibfnamefont {S.}~\bibnamefont {Imafuku}},
  \bibinfo {author} {\bibfnamefont {S.}~\bibnamefont {Kumar}}, \bibinfo
  {author} {\bibfnamefont {K.}~\bibnamefont {Nakajima}}, \bibinfo {author}
  {\bibfnamefont {M.}~\bibnamefont {Sakakura}}, \ and\ \bibinfo {author}
  {\bibfnamefont {M.}~\bibnamefont {Suzuki}},\ }\bibfield  {title} {\enquote
  {\bibinfo {title} {Direction control of quasi-stokeslet induced by
  thermoplasmonic heating of a water vapor microbubble},}\ }\href {\doibase
  10.1038/s41598-019-41255-5} {\bibfield  {journal} {\bibinfo  {journal} {Sci.
  Rep.}\ }\textbf {\bibinfo {volume} {9}},\ \bibinfo {pages} {4770} (\bibinfo
  {year} {2019})}\BibitemShut {NoStop}%
\bibitem [{\citenamefont {Namura}\ \emph {et~al.}(2022)\citenamefont {Namura},
  \citenamefont {Hanai}, \citenamefont {Kondo}, \citenamefont {Kumar},\ and\
  \citenamefont {Suzuki}}]{Namura2022}%
  \BibitemOpen
  \bibfield  {author} {\bibinfo {author} {\bibfnamefont {K.}~\bibnamefont
  {Namura}}, \bibinfo {author} {\bibfnamefont {S.}~\bibnamefont {Hanai}},
  \bibinfo {author} {\bibfnamefont {S.}~\bibnamefont {Kondo}}, \bibinfo
  {author} {\bibfnamefont {S.}~\bibnamefont {Kumar}}, \ and\ \bibinfo {author}
  {\bibfnamefont {M.}~\bibnamefont {Suzuki}},\ }\bibfield  {title} {\enquote
  {\bibinfo {title} {Gold micropetals self‐assembled by shadow‐sphere
  lithography for optofluidic control},}\ }\href {\doibase
  10.1002/admi.202200200} {\bibfield  {journal} {\bibinfo  {journal} {Adv.
  Mater. Interfaces}\ }\textbf {\bibinfo {volume} {9}},\ \bibinfo {pages}
  {2200200} (\bibinfo {year} {2022})}\BibitemShut {NoStop}%
\bibitem [{\citenamefont {Namura}\ \emph {et~al.}(2020)\citenamefont {Namura},
  \citenamefont {Okai}, \citenamefont {Kumar}, \citenamefont {Nakajima},\ and\
  \citenamefont {Suzuki}}]{Namura2020}%
  \BibitemOpen
  \bibfield  {author} {\bibinfo {author} {\bibfnamefont {K.}~\bibnamefont
  {Namura}}, \bibinfo {author} {\bibfnamefont {S.}~\bibnamefont {Okai}},
  \bibinfo {author} {\bibfnamefont {S.}~\bibnamefont {Kumar}}, \bibinfo
  {author} {\bibfnamefont {K.}~\bibnamefont {Nakajima}}, \ and\ \bibinfo
  {author} {\bibfnamefont {M.}~\bibnamefont {Suzuki}},\ }\bibfield  {title}
  {\enquote {\bibinfo {title} {Self-oscillation of locally heated water vapor
  microbubbles in degassed water},}\ }\href {\doibase 10.1002/admi.202000483}
  {\bibfield  {journal} {\bibinfo  {journal} {Adv. Mater. Interfaces}\ }\textbf
  {\bibinfo {volume} {7}},\ \bibinfo {pages} {2000483} (\bibinfo {year}
  {2020})}\BibitemShut {NoStop}%
\bibitem [{\citenamefont {Nguyen}\ \emph
  {et~al.}(2019{\natexlab{a}})\citenamefont {Nguyen}, \citenamefont
  {Sanathanan}, \citenamefont {Miao}, \citenamefont {Rivas},\ and\
  \citenamefont {Ohl}}]{Nguyen2019_2}%
  \BibitemOpen
  \bibfield  {author} {\bibinfo {author} {\bibfnamefont {D.~M.}\ \bibnamefont
  {Nguyen}}, \bibinfo {author} {\bibfnamefont {M.~S.}\ \bibnamefont
  {Sanathanan}}, \bibinfo {author} {\bibfnamefont {J.}~\bibnamefont {Miao}},
  \bibinfo {author} {\bibfnamefont {D.~F.}\ \bibnamefont {Rivas}}, \ and\
  \bibinfo {author} {\bibfnamefont {C.-D.}\ \bibnamefont {Ohl}},\ }\bibfield
  {title} {\enquote {\bibinfo {title} {In-phase synchronization between two
  auto-oscillating bubbles},}\ }\href {\doibase 10.1103/PhysRevFluids.4.043601}
  {\bibfield  {journal} {\bibinfo  {journal} {Phys. Rev. Fluid}\ }\textbf
  {\bibinfo {volume} {4}},\ \bibinfo {pages} {043601} (\bibinfo {year}
  {2019}{\natexlab{a}})}\BibitemShut {NoStop}%
\bibitem [{\citenamefont {Nguyen}\ \emph
  {et~al.}(2019{\natexlab{b}})\citenamefont {Nguyen}, \citenamefont {Supponen},
  \citenamefont {Miao}, \citenamefont {Farhat},\ and\ \citenamefont
  {Ohl}}]{Nguyen2019}%
  \BibitemOpen
  \bibfield  {author} {\bibinfo {author} {\bibfnamefont {D.~M.}\ \bibnamefont
  {Nguyen}}, \bibinfo {author} {\bibfnamefont {O.}~\bibnamefont {Supponen}},
  \bibinfo {author} {\bibfnamefont {J.}~\bibnamefont {Miao}}, \bibinfo {author}
  {\bibfnamefont {M.}~\bibnamefont {Farhat}}, \ and\ \bibinfo {author}
  {\bibfnamefont {C.-D.}\ \bibnamefont {Ohl}},\ }\bibfield  {title} {\enquote
  {\bibinfo {title} {Gravity-independent oscillate boiling},}\ }\href {\doibase
  10.1007/s12217-019-09708-8} {\bibfield  {journal} {\bibinfo  {journal}
  {Microgravity Sci. Technol.}\ }\textbf {\bibinfo {volume} {31}},\ \bibinfo
  {pages} {767--773} (\bibinfo {year} {2019}{\natexlab{b}})}\BibitemShut
  {NoStop}%
\bibitem [{\citenamefont {Wang}\ \emph {et~al.}(2017)\citenamefont {Wang},
  \citenamefont {Zaytsev}, \citenamefont {The}, \citenamefont {Eijkel},
  \citenamefont {Zandvliet}, \citenamefont {Zhang},\ and\ \citenamefont
  {Lohse}}]{WangY2017}%
  \BibitemOpen
  \bibfield  {author} {\bibinfo {author} {\bibfnamefont {Y.}~\bibnamefont
  {Wang}}, \bibinfo {author} {\bibfnamefont {M.~E.}\ \bibnamefont {Zaytsev}},
  \bibinfo {author} {\bibfnamefont {H.~L.}\ \bibnamefont {The}}, \bibinfo
  {author} {\bibfnamefont {J.~C.}\ \bibnamefont {Eijkel}}, \bibinfo {author}
  {\bibfnamefont {H.~J.}\ \bibnamefont {Zandvliet}}, \bibinfo {author}
  {\bibfnamefont {X.}~\bibnamefont {Zhang}}, \ and\ \bibinfo {author}
  {\bibfnamefont {D.}~\bibnamefont {Lohse}},\ }\bibfield  {title} {\enquote
  {\bibinfo {title} {Vapor and gas-bubble growth dynamics around
  laser-irradiated, water-immersed plasmonic nanoparticles},}\ }\href {\doibase
  10.1021/acsnano.6b08229} {\bibfield  {journal} {\bibinfo  {journal} {ACS
  Nano}\ }\textbf {\bibinfo {volume} {11}},\ \bibinfo {pages} {2045--2051}
  (\bibinfo {year} {2017})}\BibitemShut {NoStop}%
\bibitem [{\citenamefont {Wang}\ \emph {et~al.}(2018)\citenamefont {Wang},
  \citenamefont {Zaytsev}, \citenamefont {Lajoinie}, \citenamefont {The},
  \citenamefont {Eijkel}, \citenamefont {van~den Berg}, \citenamefont
  {Versluis}, \citenamefont {Weckhuysen}, \citenamefont {Zhang}, \citenamefont
  {Zandvliet},\ and\ \citenamefont {Lohse}}]{WangY2018}%
  \BibitemOpen
  \bibfield  {author} {\bibinfo {author} {\bibfnamefont {Y.}~\bibnamefont
  {Wang}}, \bibinfo {author} {\bibfnamefont {M.~E.}\ \bibnamefont {Zaytsev}},
  \bibinfo {author} {\bibfnamefont {G.}~\bibnamefont {Lajoinie}}, \bibinfo
  {author} {\bibfnamefont {H.~L.}\ \bibnamefont {The}}, \bibinfo {author}
  {\bibfnamefont {J.~C.~T.}\ \bibnamefont {Eijkel}}, \bibinfo {author}
  {\bibfnamefont {A.}~\bibnamefont {van~den Berg}}, \bibinfo {author}
  {\bibfnamefont {M.}~\bibnamefont {Versluis}}, \bibinfo {author}
  {\bibfnamefont {B.~M.}\ \bibnamefont {Weckhuysen}}, \bibinfo {author}
  {\bibfnamefont {X.}~\bibnamefont {Zhang}}, \bibinfo {author} {\bibfnamefont
  {H.~J.~W.}\ \bibnamefont {Zandvliet}}, \ and\ \bibinfo {author}
  {\bibfnamefont {D.}~\bibnamefont {Lohse}},\ }\bibfield  {title} {\enquote
  {\bibinfo {title} {Giant and explosive plasmonic bubbles by delayed
  nucleation},}\ }\href {\doibase 10.1073/pnas.1805912115} {\bibfield
  {journal} {\bibinfo  {journal} {PNAS}\ }\textbf {\bibinfo {volume} {115}},\
  \bibinfo {pages} {7676} (\bibinfo {year} {2018})}\BibitemShut {NoStop}%
\bibitem [{\citenamefont {Li}\ \emph {et~al.}(2019)\citenamefont {Li},
  \citenamefont {Wang}, \citenamefont {Zaytsev}, \citenamefont {Lajoinie},
  \citenamefont {Le~The}, \citenamefont {Bomer}, \citenamefont {Eijkel},
  \citenamefont {Zandvliet}, \citenamefont {Zhang},\ and\ \citenamefont
  {Lohse}}]{LiXiaolai2019}%
  \BibitemOpen
  \bibfield  {author} {\bibinfo {author} {\bibfnamefont {X.}~\bibnamefont
  {Li}}, \bibinfo {author} {\bibfnamefont {Y.}~\bibnamefont {Wang}}, \bibinfo
  {author} {\bibfnamefont {M.~E.}\ \bibnamefont {Zaytsev}}, \bibinfo {author}
  {\bibfnamefont {G.}~\bibnamefont {Lajoinie}}, \bibinfo {author}
  {\bibfnamefont {H.}~\bibnamefont {Le~The}}, \bibinfo {author} {\bibfnamefont
  {J.~G.}\ \bibnamefont {Bomer}}, \bibinfo {author} {\bibfnamefont {J.~C.~T.}\
  \bibnamefont {Eijkel}}, \bibinfo {author} {\bibfnamefont {H.~J.~W.}\
  \bibnamefont {Zandvliet}}, \bibinfo {author} {\bibfnamefont {X.}~\bibnamefont
  {Zhang}}, \ and\ \bibinfo {author} {\bibfnamefont {D.}~\bibnamefont
  {Lohse}},\ }\bibfield  {title} {\enquote {\bibinfo {title} {Plasmonic bubble
  nucleation and growth in water: Effect of dissolved air},}\ }\href {\doibase
  10.1021/acs.jpcc.9b05374} {\bibfield  {journal} {\bibinfo  {journal} {J.
  Phys. Chem. C}\ }\textbf {\bibinfo {volume} {123}},\ \bibinfo {pages}
  {23586--23593} (\bibinfo {year} {2019})}\BibitemShut {NoStop}%
\bibitem [{\citenamefont {Zaytsev}\ \emph {et~al.}(2020)\citenamefont
  {Zaytsev}, \citenamefont {Wang}, \citenamefont {Zhang}, \citenamefont
  {Lajoinie}, \citenamefont {Zhang}, \citenamefont {Prosperetti}, \citenamefont
  {Zandvliet},\ and\ \citenamefont {Lohse}}]{Zaytsev2020}%
  \BibitemOpen
  \bibfield  {author} {\bibinfo {author} {\bibfnamefont {M.~E.}\ \bibnamefont
  {Zaytsev}}, \bibinfo {author} {\bibfnamefont {Y.}~\bibnamefont {Wang}},
  \bibinfo {author} {\bibfnamefont {Y.}~\bibnamefont {Zhang}}, \bibinfo
  {author} {\bibfnamefont {G.}~\bibnamefont {Lajoinie}}, \bibinfo {author}
  {\bibfnamefont {X.}~\bibnamefont {Zhang}}, \bibinfo {author} {\bibfnamefont
  {A.}~\bibnamefont {Prosperetti}}, \bibinfo {author} {\bibfnamefont
  {H.~J.~W.}\ \bibnamefont {Zandvliet}}, \ and\ \bibinfo {author}
  {\bibfnamefont {D.}~\bibnamefont {Lohse}},\ }\bibfield  {title} {\enquote
  {\bibinfo {title} {Gas^^e2^^80^^93vapor interplay in plasmonic bubble
  shrinkage},}\ }\href {\doibase 10.1021/acs.jpcc.9b10675} {\bibfield
  {journal} {\bibinfo  {journal} {J. Phys. Chem. C}\ }\textbf {\bibinfo
  {volume} {124}},\ \bibinfo {pages} {5861--5869} (\bibinfo {year}
  {2020})}\BibitemShut {NoStop}%
\bibitem [{\citenamefont {Sarabia-Alonso}\ \emph {et~al.}(2020)\citenamefont
  {Sarabia-Alonso}, \citenamefont {Ortega-Mendoza}, \citenamefont
  {Ram^^c3^^adrez-San-Juan}, \citenamefont {Zaca-Mor^^c3^^a1n}, \citenamefont
  {Ram^^c3^^adrez-Ram^^c3^^adrez}, \citenamefont {Padilla-Vivanco},
  \citenamefont {Mu^^c3^^b1oz-P^^c3^^a9rez},\ and\ \citenamefont
  {Ramos-Garc^^c3^^ada}}]{Sarabia2020}%
  \BibitemOpen
  \bibfield  {author} {\bibinfo {author} {\bibfnamefont {J.~A.}\ \bibnamefont
  {Sarabia-Alonso}}, \bibinfo {author} {\bibfnamefont {J.~G.}\ \bibnamefont
  {Ortega-Mendoza}}, \bibinfo {author} {\bibfnamefont {J.~C.}\ \bibnamefont
  {Ram^^c3^^adrez-San-Juan}}, \bibinfo {author} {\bibfnamefont
  {P.}~\bibnamefont {Zaca-Mor^^c3^^a1n}}, \bibinfo {author} {\bibfnamefont
  {J.}~\bibnamefont {Ram^^c3^^adrez-Ram^^c3^^adrez}}, \bibinfo {author}
  {\bibfnamefont {A.}~\bibnamefont {Padilla-Vivanco}}, \bibinfo {author}
  {\bibfnamefont {F.~M.}\ \bibnamefont {Mu^^c3^^b1oz-P^^c3^^a9rez}}, \ and\
  \bibinfo {author} {\bibfnamefont {R.}~\bibnamefont {Ramos-Garc^^c3^^ada}},\
  }\bibfield  {title} {\enquote {\bibinfo {title} {Optothermal generation,
  trapping, and manipulation of microbubbles},}\ }\href {\doibase
  10.1364/OE.389980} {\bibfield  {journal} {\bibinfo  {journal} {Opt. Express}\
  }\textbf {\bibinfo {volume} {28}},\ \bibinfo {pages} {17672--17682} (\bibinfo
  {year} {2020})}\BibitemShut {NoStop}%
\bibitem [{\citenamefont {Detert}\ \emph {et~al.}(2020)\citenamefont {Detert},
  \citenamefont {Zeng}, \citenamefont {Wang}, \citenamefont {Le~The},
  \citenamefont {Zandvliet},\ and\ \citenamefont {Lohse}}]{Detert2020}%
  \BibitemOpen
  \bibfield  {author} {\bibinfo {author} {\bibfnamefont {M.}~\bibnamefont
  {Detert}}, \bibinfo {author} {\bibfnamefont {B.}~\bibnamefont {Zeng}},
  \bibinfo {author} {\bibfnamefont {Y.}~\bibnamefont {Wang}}, \bibinfo {author}
  {\bibfnamefont {H.}~\bibnamefont {Le~The}}, \bibinfo {author} {\bibfnamefont
  {H.~J.~W.}\ \bibnamefont {Zandvliet}}, \ and\ \bibinfo {author}
  {\bibfnamefont {D.}~\bibnamefont {Lohse}},\ }\bibfield  {title} {\enquote
  {\bibinfo {title} {Plasmonic bubble nucleation in binary liquids},}\ }\href
  {\doibase 10.1021/acs.jpcc.9b10064} {\bibfield  {journal} {\bibinfo
  {journal} {J. Phys. Chem. C Nanomater. Interfaces}\ }\textbf {\bibinfo
  {volume} {124}},\ \bibinfo {pages} {2591--2597} (\bibinfo {year}
  {2020})}\BibitemShut {NoStop}%
\bibitem [{\citenamefont {Baffou}\ \emph {et~al.}(2014)\citenamefont {Baffou},
  \citenamefont {Polleux}, \citenamefont {Rigneault},\ and\ \citenamefont
  {Monneret}}]{Baffou2014}%
  \BibitemOpen
  \bibfield  {author} {\bibinfo {author} {\bibfnamefont {G.}~\bibnamefont
  {Baffou}}, \bibinfo {author} {\bibfnamefont {J.}~\bibnamefont {Polleux}},
  \bibinfo {author} {\bibfnamefont {H.}~\bibnamefont {Rigneault}}, \ and\
  \bibinfo {author} {\bibfnamefont {S.}~\bibnamefont {Monneret}},\ }\bibfield
  {title} {\enquote {\bibinfo {title} {Super-heating and micro-bubble
  generation around plasmonic nanoparticles under cw illumination},}\ }\href
  {\doibase 10.1021/jp411519k} {\bibfield  {journal} {\bibinfo  {journal} {J.
  Phys. Chem. C}\ }\textbf {\bibinfo {volume} {118}},\ \bibinfo {pages}
  {4890--4898} (\bibinfo {year} {2014})}\BibitemShut {NoStop}%
\bibitem [{\citenamefont {Tomiyama}\ \emph {et~al.}(1998)\citenamefont
  {Tomiyama}, \citenamefont {Kataoka}, \citenamefont {Zun},\ and\ \citenamefont
  {Sakaguchi}}]{Tomiyama1998}%
  \BibitemOpen
  \bibfield  {author} {\bibinfo {author} {\bibfnamefont {A.}~\bibnamefont
  {Tomiyama}}, \bibinfo {author} {\bibfnamefont {I.}~\bibnamefont {Kataoka}},
  \bibinfo {author} {\bibfnamefont {I.}~\bibnamefont {Zun}}, \ and\ \bibinfo
  {author} {\bibfnamefont {T.}~\bibnamefont {Sakaguchi}},\ }\bibfield  {title}
  {\enquote {\bibinfo {title} {Drag coefficients of single bubbles under normal
  and micro gravity conditions},}\ }\href {\doibase 10.1299/jsmeb.41.472}
  {\bibfield  {journal} {\bibinfo  {journal} {JSME Int. J. Ser. B}\ }\textbf
  {\bibinfo {volume} {41}},\ \bibinfo {pages} {472--479} (\bibinfo {year}
  {1998})}\BibitemShut {NoStop}%
\end{thebibliography}%

\end{document}